\documentclass{aa}  

\usepackage{pifont}
\usepackage{amsmath}
\usepackage{graphicx}
\usepackage{txfonts}
\usepackage{url}
\usepackage{longtable}
\usepackage{lscape}
\usepackage{natbib}
\usepackage[colorlinks,linkcolor=red,anchorcolor=green,citecolor=blue]{hyperref}

\begin{document}

   \title{Gas kinematics and star formation in the filamentary molecular cloud G47.06+0.26 }
  \author{Jin-Long Xu,
          \inst{1,5}
          Ye Xu,\inst{2,6}
          Chuan-Peng Zhang,\inst{1,5}
          Xiao-Lan Liu,\inst{1,5}
          Naiping Yu,\inst{1,5}
          Chang-Chun Ning,\inst{3,5}
          \ and Bing-Gang Ju\inst{4,6}
          }
   \institute{National Astronomical Observatories, Chinese Academy of Sciences,
             Beijing 100012, China \\
         \email{xujl@bao.ac.cn}
          \and
          Purple Mountain Observatory, Chinese Academy of Sciences, Nanjing 210008, China
          \and
          Tibet University, Lhasa,  Tibet 850000, China
          \and
          Purple Mountain Observatory, Qinghai Station,Delingha 817000, China
         \and
          NAOC-TU Joint Center for Astrophysics, Lhasa 850000, China
          \and Key Laboratory of Radio Astronomy, Chinese Academy of Sciences, China
        }
\authorrunning{J.-L. Xu et. al.}
\titlerunning{Gas kinematics and Star Formation in G47.06+0.26}

   \date{Received XXX, XXX; accepted XXX, XXX}

\abstract
   {}
{We performed a multi-wavelength study toward the filamentary cloud G47.06+0.26 to investigate the gas kinematics and star formation.}
{We present the $^{12}$CO ($J$=1-0), $^{13}$CO ($J$=1-0) and C$^{18}$O ($J$=1-0) observations of G47.06+0.26 obtained with the Purple Mountain Observation (PMO) 13.7 m radio telescope to investigate the detailed kinematics of the filament. Radio continuum and infrared archival data were obtained from the NRAO VLA Sky Survey  (NVSS),  the  APEX Telescope Large Area Survey of the Galaxy (ATLASGAL), the Galactic Legacy Infrared Mid-Plane Survey Extraordinaire (GLIMPSE) survey, and the Multi-band Imaging Photometer Survey of the Galaxy (MIPSGAL). To trace massive clumps and extract YSOs in G47.06+0.26, we used the BGPS catalog v2.0 and the GLIMPSE I catalog, respectively.}
{The $^{12}$CO ($J$=1-0) and $^{13}$CO ($J$=1-0) emission of G47.06+0.26 appear to show a filamentary structure.
The filament extends about 45$^{\prime}$ (58.1 pc) along the east-west direction. The mean width is about 6.8 pc, as traced by the $^{13}$CO ($J$=1-0) emission. G47.06+0.26  has a linear mass density of $\sim$361.5 $M_{\odot}$ $\rm pc^{-1}$. The external pressure (due to  neighboring bubbles and H {\small II} regions) may help preventing the filament from dispersing under the effects of turbulence. From the velocity-field map, we discern a velocity gradient perpendicular to G47.06+0.26. From the Bolocam Galactic Plane Survey (BGPS) catalog, we found nine BGPS sources in G47.06+0.26, that appear to these sources have sufficient mass to form massive stars. We obtained that the clump formation efficiency (CFE) is  $\sim$18$\%$ in the filament. Four infrared bubbles were found to be located in, and adjacent to,  G47.06+0.26. Particularly, infrared bubble N98 shows a cometary structure. CO molecular gas adjacent to N98 also shows a very intense emission. H {\small II} regions associated with infrared bubbles can inject the energy to surrounding gas. We calculated the kinetic energy, ionization energy, and thermal energy of two H {\small II} regions in  G47.06+0.26. From the GLIMPSE I catalog, we selected some Class I sources with an age of $\sim10^{5}$ yr, which are clustered along the filament. The feedback from the H {\small II} regions may cause the formation of a new generation of stars in filament G47.06+0.26.}
   {}


   \maketitle
%

\section{Introduction}
Massive stars have a significant effect on the Galactic environment, but their formation is not well understood. It is known that 70\%-90\% of stars are born in clusters in giant molecular clouds \citep[GMCs,][]{Lada2003}, which usually show filamentary structures \citep{Busquet2013}. Filaments have recently attracted  a lot of attention, especially as some dust continuum Galactic plane surveys have revealed a wealth of filamentary structures in star-forming clouds, such as the APEX Telescope Large Area Survey of the Galaxy (ATLASGAL) \citep{Schuller2009} and the Herschel infrared Galactic Plane Survey (Hi-GAL) \citep{Molinari}. Massive stars predominantly form in the most bound parts of filaments  \citep{Andre2014}, while low-mass protostars tend to form in less-bound sites \citep{Foster,Xu}.

Theoretical works show that filamentary molecular clouds can fragment into clumps due to gravitational instabilities \citep{Inutsuka,Nakamura,Tomisaka}. These clumps are not randomly distributed in the filamentary molecular cloud, but along filaments, like beads on a string. Once the clumps are formed, some of them will evolve, collapse, and form stars. \citet{Contreras} found a total of 101 clumps within five filaments using the ATLASGAL 870 $\mu$m continuum data. They found that most clumps have sufficient mass and density to form massive stars. Moreover, the filaments can also enhance accretion rates of individual star-forming cores by material flowing from the filament to the clumps \citep{Banerjee,Myers}. Hence, filamentary molecular clouds are usually associated with massive star-forming regions. 
Once a massive star forms inside a filament, UV radiation and stellar winds will ionize the surrounding gas and create an infrared bubble \citep{Churchwell,zhang}, and perhaps even disrupt the natal molecular cloud \citep{Jackson}. In addition, the expanding H {\small II} region and stellar winds will entrain and accelerate ambient gas and inject momentum and energy into the surrounding environment \citep{Narayanan,Arce}. H {\small II} regions contain kinetic energy, ionization energy and thermal energy.  Hence,  feedback from young massive stars has been proposed as a significant aspect of the self-regulation of star formation \citep[e.g.,][]{Norman,Franco,Liu,Dewangan}.  Particularly, if feedback can maintain the observed turbulence in molecular clouds, then it can be responsible for stabilizing the clouds against gravitational collapse \citep{Shu}.

GMC G47.06+0.26, with a total mass of 2.0$\times$10$^{4}$ M$_{\odot}$ and a distance of 4.44 kpc, shows a filamentary structure \citep{Wang}.  G47.06+0.26, as a long, dense GMC, is a good laboratory to investigate the connection between filaments and star formation, and to understand the processes that lead to the fragmentation of a filament.

In this paper, we present the results of a multi-wavelength study to investigate the gas kinematics and star formation in filament G47.06+0.26. Combining our data with those obtained by the NRAO VLA Sky survey, the ATLASGAL survey, and the GLIMPSE survey, our aim was to construct a comprehensive large-scale picture of G47.06+0.26. Our observations and data reduction are described in Sect.2, and the results are presented in Sect.3. In Sect.4, we  discuss the star-formation scenario, while our conclusions are summarized in Sect.5.

\section{Observations and data reduction}
The mapping observations of filament G47.06+0.26 were performed in the $^{12}$CO
($J$=1-0), $^{13}$CO ($J$=1-0), and C$^{18}$O ($J$=1-0) lines using the Purple Mountain Observation (PMO) 13.7 m radio telescope  at De Ling Ha in western China at an altitude of 3200 meters, during October 2015.  The 3$\times$3 beam array receiver system in single-sideband (SSB) mode was used as front end. The back end is a fast Fourier transform spectrometer (FFTS) of 16384 channels with a bandwidth of 1 GHz. $^{12}$CO $J$=1-0 was observed at upper sideband with a system noise temperature (Tsys) of 270 K, while $^{13}$CO $J$=1-0 and C$^{18}$O $J$=1-0 were observed simultaneously at lower sideband with a  system noise temperature of 146 K. The half-power beam width (HPBW) was 53$^{\prime\prime}$ at 115 GHz. The total mapping area is $30^{\prime}\times 20^{\prime}$ in $^{12}$CO ($J$=1-0), $^{13}$CO ($J$=1-0), and C$^{18}$O ($J$=1-0) with a $0.5^{\prime}\times0.5^{\prime}$ grid. \citet{Xu}  gave more details of the PMO 13.7 m radio telescope instrumentation.

In addition, we also used survey data from different wavebands, which are listed in Table 1.
\begin{table}[]
\begin{center}
\tabcolsep 1.2mm \caption{ The different surveys}
\def\temptablewidth{10\textwidth}
\begin{tabular}{lccc}
\hline\hline\noalign{\smallskip}
Survey  &  Waveband & Reference \\
  \hline\noalign{\smallskip}
NRAO VLA Sky&   21 cm&     \citep{Condon1998}  \\
BGPS        &  1.1 mm &    \cite{Ginsburg}   \\
ATLASGAL&     870 $\mu$m&  \citep{Schuller2009}  \\
GLIMPSE&    3.6, 4.5, 5.8, 8.0 $\mu$m&   \citep{Benjamin03} \\
MIPSGAL&   24 $\mu$m&    \citep{Carey} \\
\noalign{\smallskip}\hline
\end{tabular}\end{center}
\end{table}

\section{Results}
\subsection{Carbon monoxide, infrared, and radio continuum emission of G47.06+0.26}

Figure 1 shows the averaged spectra of $^{12}$CO ($J$=1-0), $^{13}$CO ($J$=1-0), and C$^{18}$O ($J$=1-0) over the filamentary molecular cloud G47.06+0.26. We find that there are two main components at velocity intervals 30$\sim$50 km s$^{-1}$ and 50$\sim$67 km s$^{-1}$, respectively. The LSR velocity of G47.06+0.26 is 57.5 km s$^{-1}$ in $^{13}$CO ($J$=1-0) \citep{Wang}, then the CO emission component in the velocity interval of 50$\sim$67 km s$^{-1}$ is associated with G47.06+0.26. The C$^{18}$O ($J$=1-0) emission of G47.06+0.26 is very weak (0.13 K). The $^{12}$CO ($J$=1-0) emission shows another component at 30$\sim$50 km s$^{-1}$, originating from independent material along the line of sight. Using the Herschel Infrared Galactic Plane Survey data, \citet{Wang} found that the dust continuum emission of G47.06+0.26 shows a bent C-shape. To find any associated components with  G47.06+0.26, we made channel maps in $^{13}$CO ($J$=1-0). Through the channel maps, we obtain that the $^{13}$CO ($J$=1-0) emission detected in the velocity range between 52 and 64 km s$^{-1}$ are well correlated with the dust continuum emission of G47.06+0.26. The velocity component in the interval 30$\sim$50 km s$^{-1}$  is likely foreground or background emission not related with G47.06+0.26. 

\begin{figure}[]
\vspace{-5mm}
\includegraphics[angle=270,scale=.39]{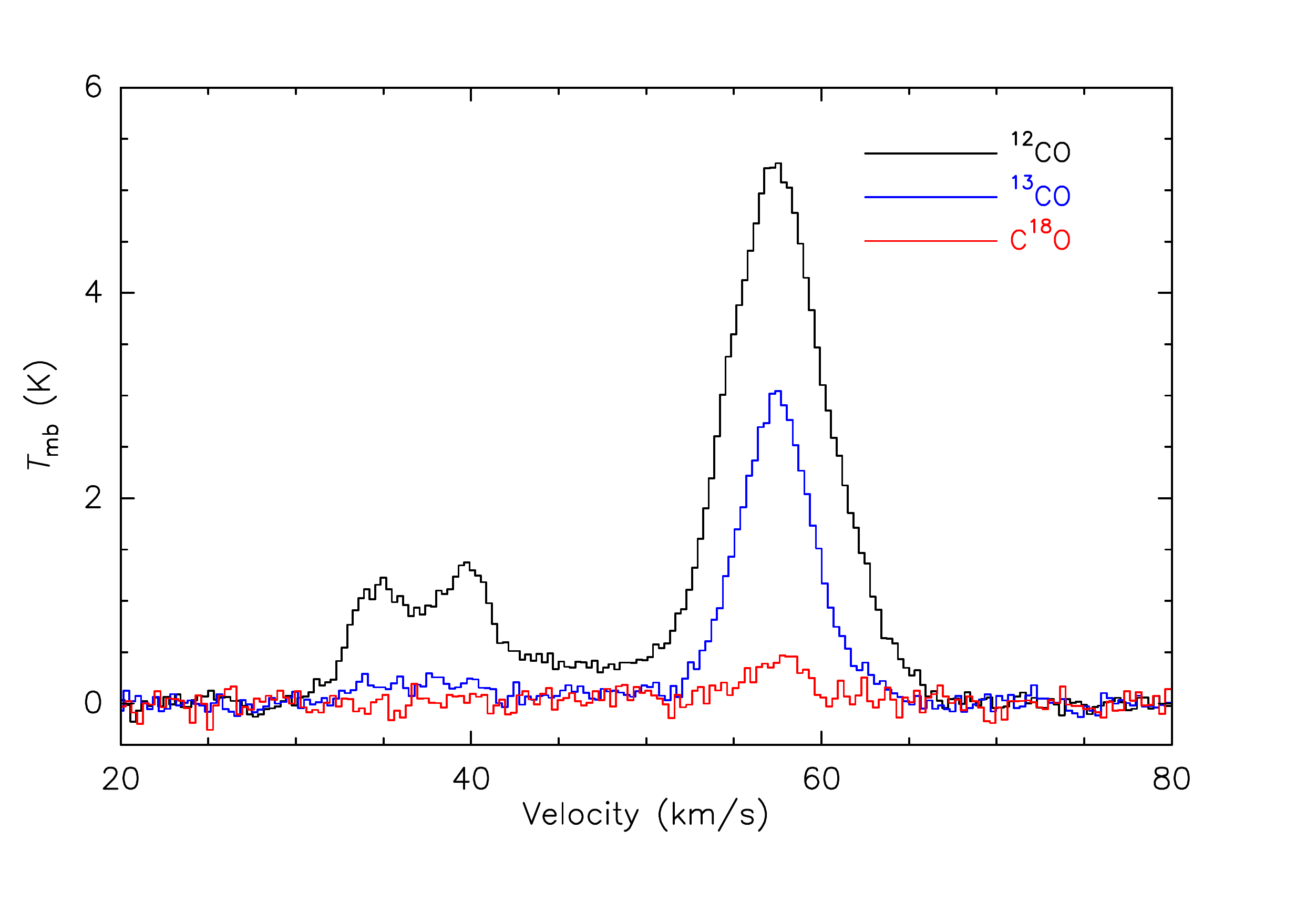}
\vspace{-8mm}\caption{Averaged spectra of $^{12}$CO ($J$=1-0), $^{13}$CO ($J$=1-0), and C$^{18}$O ($J$=1-0)  over the entire filament. We note that the intensities of $^{13}$CO ($J$=1-0) and C$^{18}$O ($J$=1-0) are multiplied by factors of two and three for clarity, respectively. }
\end{figure}

\begin{figure*}[]
\vspace{-42mm}
\centering
\includegraphics[angle=0,scale=0.79]{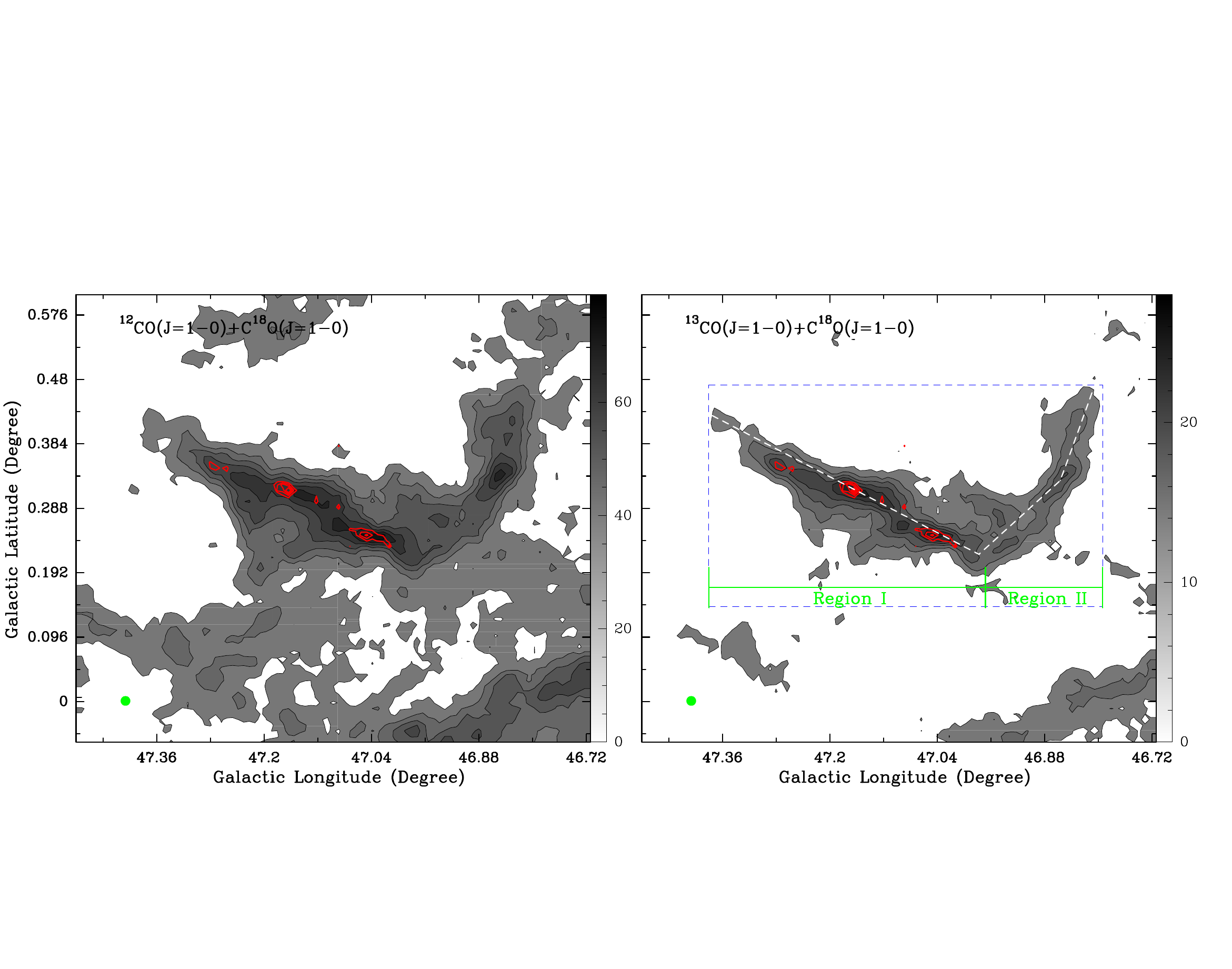}
\vspace{-30mm}\caption{Left panel: $^{12}$CO ($J$=1-0) emission map (gray scale) overlaid with C$^{18}$O ($J$=1-0) emission contours (red). The gray  levels range from 11.0 (5$\sigma$) to 77.0  in steps of 11.0 K km s$^{-1}$, while the red contour levels range  from 3.0 (5$\sigma$) to 4.8 in steps of 0.6 K km s$^{-1}$. Right panel: $^{13}$CO ($J$=1-0) integrated emission map. The  gray levels range from 3.0 (5$\sigma$) to 24.0 in steps of 4.2 K km s$^{-1}$. In both maps, the emission has been integrated from 52 to 64 km s$^{-1}$. The dashed box indicates our studied filament. The white dashed lines indicate the direction of position-velocity diagrams (See Figure 9). The beam size is shown in the lower-left corner. The scale  units in each panel of both axes are in degrees.}
\end{figure*}

Using the velocity range of 52 to 64 km s$^{-1}$, we made integrated intensity maps of $^{12}$CO ($J$=1-0), $^{13}$CO ($J$=1-0), and C$^{18}$O ($J$=1-0), as shown in Fig. 2. From a large-scale perspective, the $^{12}$CO ($J$=1-0) and $^{13}$CO ($J$=1-0) emission of G47.06+0.26 shows a filamentary structure extending from west to east. The region traced by $^{12}$CO ($J$=1-0) is more diffuse than that of $^{13}$CO ($J$=1-0). We divided the filament into region I and region II, and found that C$^{18}$O ($J$=1-0) is only seen in region I, indicating that it is  denser than in region II. Several clumps are found in the filament. Adopting a distance of 4.44 kpc \citep{Wang}, the filament extends about 58.1 pc (45.0$^{\prime}$) in length, while the mean width is about 6.8 pc (5.3$^{\prime}$), as traced by the $^{13}$CO ($J$=1-0) emission. The aspect ratio is 9:1. The measured length is a lower limit, due to the unknown inclination.

\begin{figure}[]
\vspace{0mm}
\includegraphics[angle=0,scale=.45]{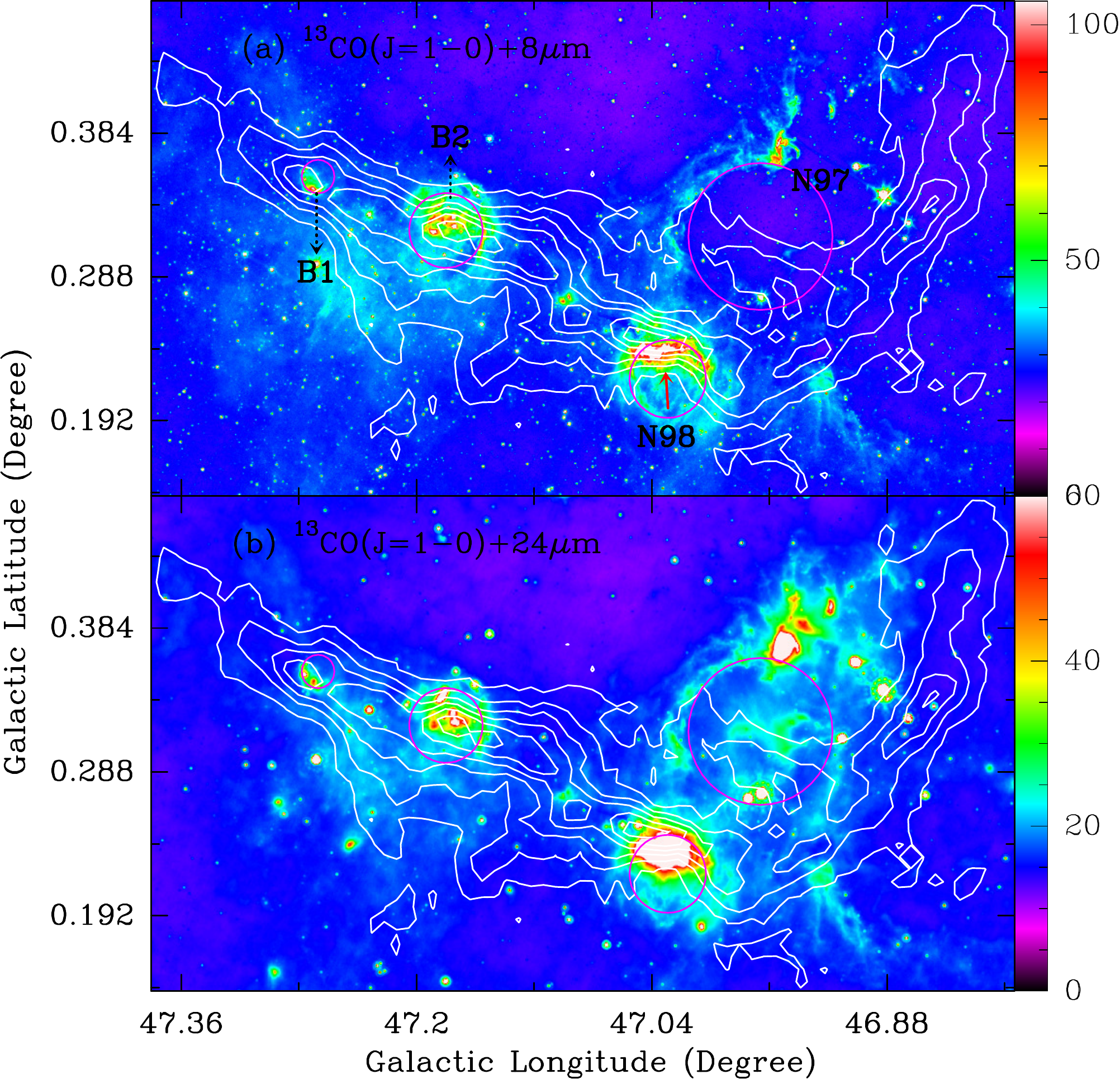}
\vspace{-4mm}\caption{(a) $^{13}$CO $J$=1-0 emission map (white contours) superimposed on the Spitzer-IRAC 8 $\mu$m emission map (color scale). The red arrow indicates steep integrated intensity gradients. (b) $^{13}$CO $J$=1-0 emission map (white contours) superimposed on the Spitzer-MIPSGAL 24 $\mu$m emission (color scale). The four pink circles represent infrared bubbles. B1 is bubble MWP1G047270+003500S, while B2 is MWP1G047180+003195. The right colour-bar is in units of MJy sr$^{-1}$. }
\end{figure}

\begin{figure*}[]
\vspace{0mm}
\includegraphics[angle=0,scale=0.72]{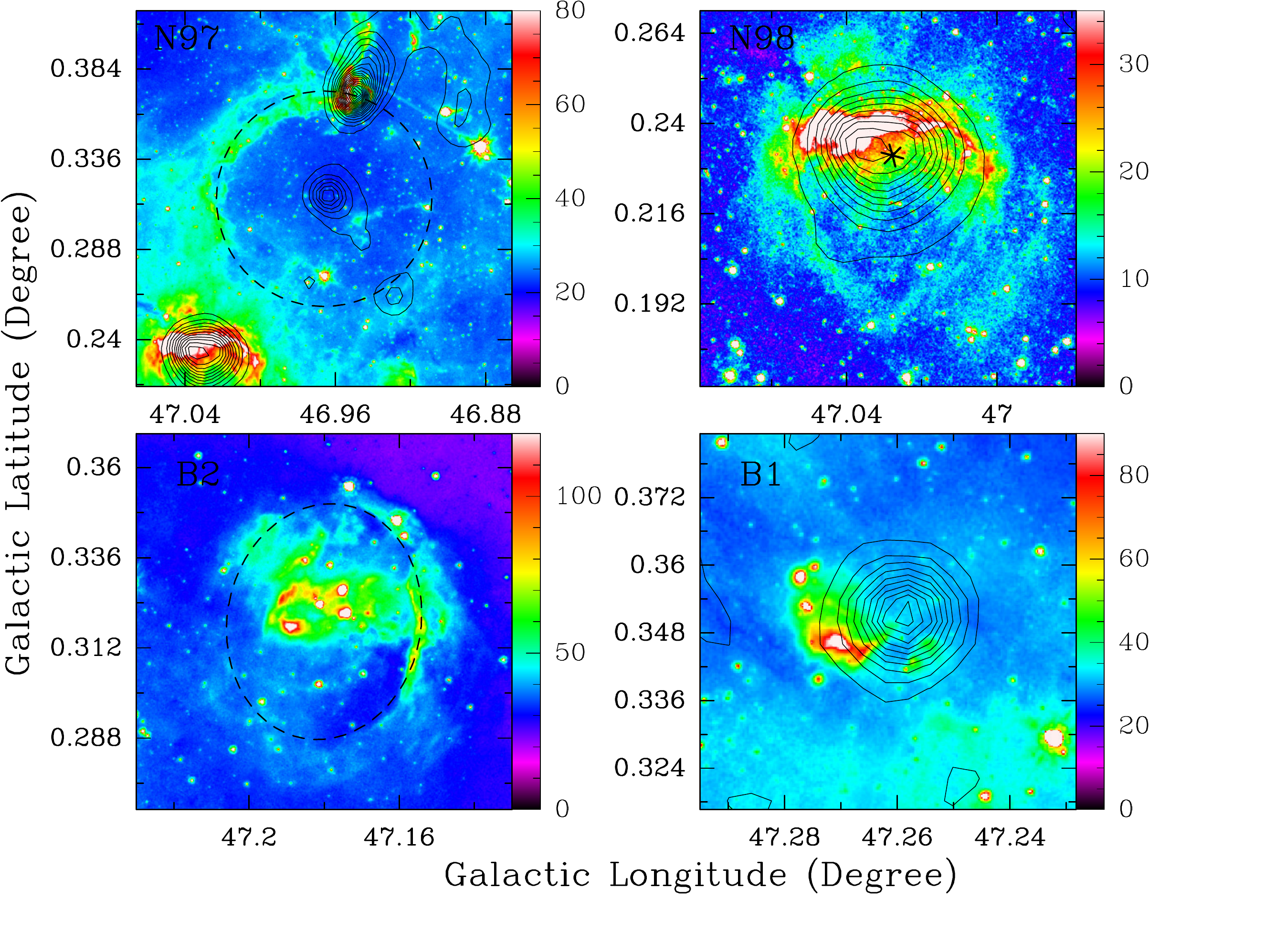}
\vspace{-12mm}\caption{The Spitzer-IRAC 8
$\mu$m emission maps (color scale) of Bubbles N97, N98, B2 and B1 overlaid with 1.4 GHz radio continuum emission
contours (black). $``\ast"$ indicates the position of  H {\small II} region G047.028+0.232 (Anderson et al. 2011). The right colour bar is in units of MJy sr$^{-1}$. }
\end{figure*}

Figure 3 (a) and (b) displays the $^{13}$CO ($J$=1-0) emission maps overlaid on the Spitzer-IRAC 8 $\mu$m and Spitzer-MIPSGAL 24 $\mu$m   emission  of G47.06+0.26, respectively. We find some bright 8 $\mu$m emission along, and adjacent to, the filamentary molecular cloud, whose image is visually similar to the 24 $\mu$m emission.  Generally, the bright Spitzer-IRAC 8 $\mu$m emission is attributed to polycyclic aromatic hydrocarbons (PAHs) \citep{Leger}. The PAH molecules are excited in the photodissociation regions (PDRs) by UV radiation arising from H {\footnotesize II} regions, however, they can also be destroyed inside the ionized gas  \citep{Pomares2009}. Hence, the Spitzer-IRAC 8 $\mu$m emission can be used to delineate an infrared bubble. Infrared bubbles are highlighted by the bright 8.0 $\mu$m emission surrounding O and early-B stars, which show full or partial ring structures  \citep{Churchwell}. \citet{Churchwell,Churchwell07} compiled a list of $\sim$600 infrared bubbles, while \citet{Simpson12} created a new catalog of 5106 infrared bubbles through visual inspection via the online citizen science website  `The Milky Way Project (MWP)'. From the above two catalogs, we identified four infrared bubbles N97, N98, MWP1G047270+003500S(B1), and  MWP1G047180+003195(B2). We use pink circles to represent the bubbles in Fig.3. N97 and N98 are close to the filamentary molecular cloud, while B1 and B2 are located within it. In order to clearly show the shape of a single bubble, we made the enlarged maps shown in Fig. 4. N97 shows an arc-like structure that is opened at the southwest, while N98 presents a cometary structure. From Fig.3, we can see that the brightest part of N98 is located within the filamentary molecular cloud.  CO molecular gas adjacent to N98 also shows an arc-like structure with a very intense emission towards N98, as shown by the red arrow in Fig. 3 (a).  The H {\small II} region G047.028+0.232 is associated with N98 \citep{Anderson11}. The RRL velocity of G047.028+0.232 is 56.9 $\pm$ 0.1 km s$^{-1}$, which is roughly consistent with the LSR velocity of the filamentary molecular cloud (57.5 km s$^{-1}$).  The   1.4 GHz radio continuum emission is mainly from free-free emission, which can be used to trace the ionized gas of H {\small II} regions. Figure 4 also presents the radio 1.4 GHz continuum emission maps overlaid on the Spitzer-IRAC 8 $\mu$m emission. The infrared bubbles N97, N98, and B1 are associated with ionized gas, while there is no 1.4 GHz radio continuum emission found for B2.

Figure 5 shows the ATLASGAL 870 $\mu$m emission of the filament overlaid with the $^{13}$CO ($J$=1-0) emission. The 870 $\mu$m emission traces the distribution of cold dust. Obviously, the $^{13}$CO $J$=1-0 emission is more extended than that of the ATLASGAL 870 $\mu$m. Comparing with the C$^{18}$O $J$=1-0 emission in Fig. 2 (right panel), the spatial distribution of 870 $\mu$m is similar to that of C$^{18}$O ($J$=1-0), which is denser adjacent to N98. Next, BGPS is a 1.1 mm survey of dust emission in the Northern Galactic plane, which has a 33$^{\prime\prime}$ beam \citep{Ginsburg}. These authors compiled a new catalog (v2.0), which includes 8594 sources. Most BGPS sources show signs of active star formation \citep{Dunham,Xi}. Using this catalog, we found  ten BGPS sources in our observed region, as shown in Fig.5. Except for one source, the remaining sources are all associated with ATLASGAL 870 $\mu$m filamentary emission  and  $^{13}$CO ($J$=1-0) emission. Table 2 lists the parameters of nine BGPS sources.

\begin{figure*}[]
\vspace{0mm}
\includegraphics[angle=0,scale=0.61]{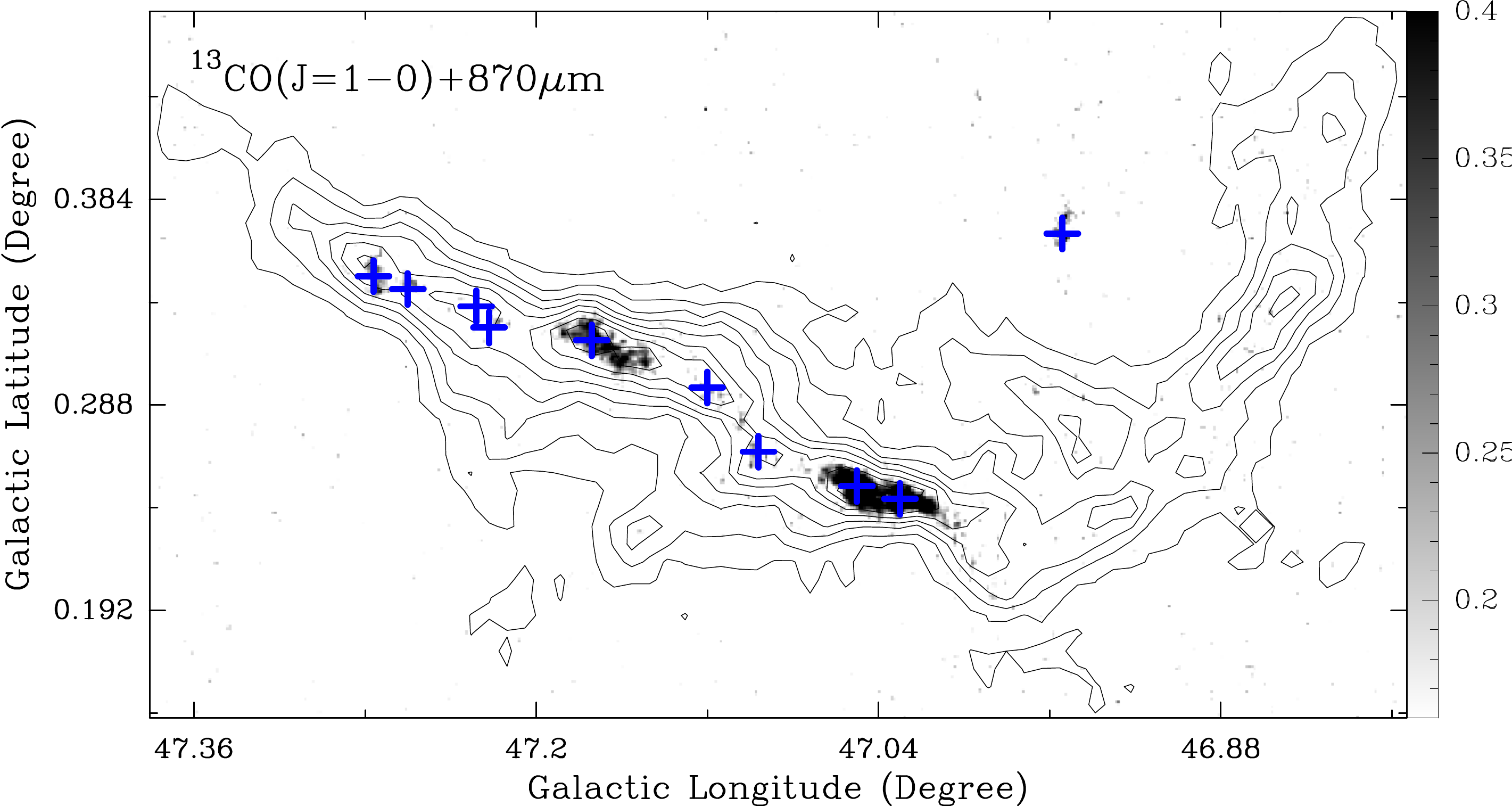}
\vspace{0mm}\caption{$^{13}$CO ($J$=1-0) integrated intensity map of the filament overlaid on the  ATLASGAL 870 $\mu$m image (gray scale). The blue pluses indicate the positions of 1.1mm BGPS sources. The right colour bar is in units of Jy beam$^{-1}$. }
\end{figure*}

\begin{figure}[]
\vspace{0mm}
\includegraphics[angle=270,scale=0.65]{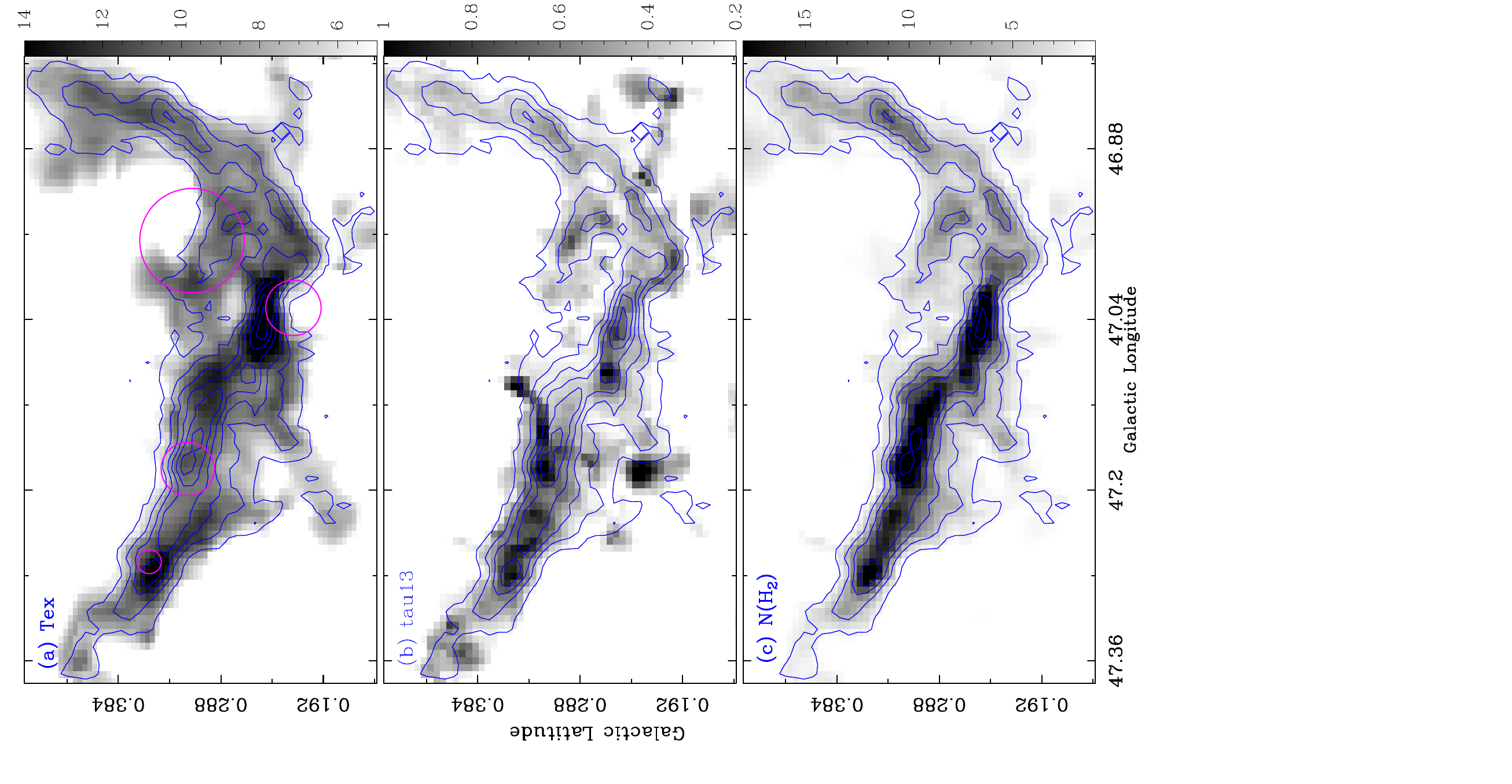}
\vspace{-43mm}\caption{(a) Map of the excitation temperature with a linear scale in units of K
from the $^{12}$CO ($J$=1-0) emission in the 52--64 km s$^{-1}$ interval. (b) Map of the optical depths of $^{13}$CO ($J$=1-0). (c) Map of the column density in units of 10$^{21}$ cm$^{-2}$. The blue contours represent the $^{13}$CO ($J$=1-0) emission map.}
\end{figure}

\begin{table*}[]
\begin{center}
\tabcolsep 4.4mm \caption{Physical parameters of the BGPS clumps. The first column is the name of the clumps. The second and third columns list the galactic longitude and latitude of the clumps, respectively. The fourth, fifth, and sixth columns list the angular sizes of major axis and minor axis, and  the position angle of the clumps, respectively. The seventh, eighth, and ninth  columns list the integrated flux density, mass, and number density of each clump, respectively.}
\def\temptablewidth{10\textwidth}
\begin{tabular}{lccccccccc}
\hline\hline\noalign{\smallskip}
Name  & $l$ & $b$ & $L_{\rm maj}$ & $L_{\rm min}$ & pos&  Flux& $M$   &$n(\rm H_{2})$ \\
      &(deg)& (deg)& (arcsec)  &(arcsec)  &(deg) &(Jy) &(M$_{\odot}$)&($10^{4}$ cm$^{-3}$)   \\
  \hline\noalign{\smallskip}
G047.030+00.244&    47.030&    0.244 &      43.62&   17.72&      64&  2.2&        568.2&  7.5  \\
G047.050+00.250&    47.050&    0.250 &      42.28&   26.09&      68&  3.7&        955.5&  7.4 \\
G047.096+00.266&    47.096&    0.266 &      37.08&   18.91&      34&  0.6&        155.0&  2.4 \\
G047.119+00.296&    47.120&    0.296 &      33.85&   18.64&      37&  0.7&        180.8&  3.2\\
G047.173+00.318&    47.174&    0.318 &      57.49&   28.75&      80&  4.0&        1033.0& 4.4 \\
G047.222+00.324&    47.222&    0.324 &      32.87&   19.84&     117&  0.8&        206.6&  3.5  \\
G047.228+00.334&    47.228&    0.334 &      19.12&   17.82&     170&  0.5&         129.1& 5.8 \\
G047.259+00.342&    47.260&    0.342 &      38.77&   18.62&      16&  0.9&         232.4& 3.4 \\
G047.276+00.348&    47.276&    0.348 &      28.80&   22.41&      43&  1.1&         284.1& 4.9 \\
\noalign{\smallskip}\hline
\end{tabular}\end{center}
\end{table*}

\subsection{Physical parameters}
\subsubsection{The column density and mass of G47.06+0.26}
To determine the column density and mass of the filamentary molecular cloud G47.06+0.26, we used the optically thin $^{13}$CO ($J$=1-0) emission. Assuming local thermodynamical equilibrium (LTE), the column density was estimated via Garden et al. \citep{Garden}
\begin{equation}
\mathit{N(\rm ^{13}CO)}=4.71\times10^{13}\frac{T_{\rm ex}+0.88}{\rm exp(-5.29/T_{\rm ex})}\frac{\tau}{\rm 1-exp(-\tau)}W ~\rm cm^{-2},
\end{equation}
where $T_{\rm ex}$ is the mean excitation temperature of the molecular gas and $\tau$ is the  optical depth. W is $\int T_{\rm mb}dv$ in units of K km s$^{-1}$, $dv$ is the velocity range and $T_{\rm mb}$ is the corrected main-beam temperature of $^{13}$CO ($J$=1-0).

Generally, the $^{12}$CO  emission is optical thick, so we used $^{12}$CO  to estimate $T_{\rm ex}$ via following the equation  \citep{Garden}
\begin{equation}
\mathit{T_{\rm ex}}=\frac{5.53}{{\ln[1+5.53/(T_{\rm mb}+0.82)]}},
\end{equation}
where $T_{\rm mb}$ is the corrected main-beam brightness temperature of $^{12}$CO. Using all emission that is greater than 5 $\sigma$, we made a map of the excitation temperature of $^{12}$CO ($J$=1-0). In Figure 6 (a), we find that the excitation
temperature is 4.4-17.0 K, while the mean excitation temperature of the filament is about 9.3 K.

Moreover, we assumed that the excitation temperature of $^{12}$CO and $^{13}$CO have the same value in the filamentary cloud.  The optical depth ($\tau$) can be derived using the following equation \citep{Garden}
\begin{equation}
\mathit{\tau(\rm ^{13}CO)}=-\ln[1-\frac{T_{\rm mb}}{5.29/[\rm exp(5.29/\it T_{\rm ex})\rm -1]-0.89}],
\end{equation}
From the above equation, we made a map of the optical depths of $^{13}$CO ($J$=1-0) (the b panel in Figure 6). We derive that the optical depth is 0.1-1.0, suggesting that the $^{13}$CO emission is optically thin in the filament.

Next we used the relation $N(\rm H_{2})/\it N(\rm ^{13}CO)$ $\approx$
$7\times10^{5}$  \citep{Castets} to estimate the H$_{2}$ column density. We also made a map of the column density, as shown in Figure 6c. The obtained column density is (0.8-30.3)$\times10^{21}$ cm$^{-2}$, while the mean column density is 6.2$\times10^{21}$ cm$^{-2}$.  Generally, filaments show approximately long cylindrical structures \citep{Jackson},  thus the mass of the filament can be given by
\begin{equation}
\mathit{M_{\rm H_{2}}}=\pi
(\frac{d}{2})^{2}l\mu_{g}m(\rm H_{2})\it n(\rm H_{2}),
\end{equation}
where $\mu_{g}$=1.36 is the mean atomic weight of the gas, $l$ is the length of the filament,  $m(\rm
H_{2})$ is the mass of a hydrogen molecule,  and $d$ is the mean width (6.8 pc) of the filament. The mean number density of $\rm H_{2}$ is estimated to be
$n(\rm H_{2})=8.1\times10^{-20}$$N(\rm H_{2})$$/d$ \citep{Garden}.  Hence, we obtained that the mean number density of the filament is 147.2 cm$^{-3}$. Using the mean number density, we derived a mass of  $\sim$2.1$\times10^{4}\rm M_{ \odot}$, which is approximatively equal to that obtained from \citet{Wang} for G47.06+0.26.

\subsubsection{The masses of BGPS clumps}
The dust continuum emission is expected to be optically thin at millimeter wavelengths in BGPS clumps  \citep{Bally}. The masses of the BGPS clumps, as derived from their dust continuum emission, were determined using the relation
\begin{equation} \mathit{M_{\rm clump}}=\frac{S_{v}D^{2}}{\kappa_{\nu}B_{\nu}(T_{d})}
\end{equation}
where $S_{v}$ is the flux density at the frequency $\nu$ and $D$ is the distance to the clumps. 
$\kappa_{\nu}$ is the dust opacity, which is adopted as 0.0114 cm$^{2}$ g$^{-1}$  \citep{Enoch}. Here the ratio of gas  to dust was taken as 100, which is included in $\kappa_{\nu}$. $B_{v}(T_{\rm d})$ is the Planck function for the dust temperature $T_{d}$ and frequency $\nu$. According to \citet{Rosolowsky}, we also used a dust temperature of 20 K for each BGPS clump. Assuming that the clumps have roughly spherical shapes, we computed the H$_{2}$ volume density via $n(\rm H_{2})=\it 3M_{\rm clump}/(4\pi r^{3}\mu_{g}m(\rm H_{2}$)), where $r$ is the effective radius of each clump, which can be determined by $r=\sqrt{L_{maj}\cdot L_{min}}$/2. Here, $L_{maj}$ and $L_{min}$ are the de-convolved major and minor axes of each clump, which are listed in Table 2. The obtained masses and volume densities of these BGPS clumps are listed in  Table 2. From Table 2, we found that these clumps have masses ranging from 129 to 1033 M$_{\odot}$ with a total mass of clumps of 3.8 $\times10^{3}\rm M_{ \odot}$.

\subsubsection{The ionizing luminosity and ages of the H {\small II} regions}
Four infrared bubbles are found to be close to the filament.  Infrared bubbles N98 and B1 are filled with ionized gas, which is traced by the 1.4 GHz radio continuum emission, suggesting there are two H {\small II} regions located within the centers of these two bubbles. Only a small region of ionized gas is detected at the  center of N97. There is no detected ionized gas in B2. Thus, we only calculated the dynamical ages of H {\small II} regions associated with  N98 and B1. Using the 1.4 GHz radio continuum emission, the ionizing luminosity $N_{\rm Ly}$ was computed via \citep{Condon92}
\begin{equation}
 \mathit{N_{\rm Ly}}=7.54\times10^{46}(\frac{\nu}{\rm GHz})^{0.1}(\frac{T_{4}}{\rm K})^{-0.45}(\frac{S_{\nu}}{\rm Jy})(\frac{D}{\rm kpc})^{2}\rm ~s^{-1},
\end{equation}
where $\nu$ is the frequency of the observation, $T_{4}$ is the effective electron temperature in units of 10$^{4}$ K, $S_{\nu}$ is the observed specific flux density, and $D$ is the distance to the H {\small II} region. Here we adopted an effective electron temperature of 10$^{4}$ K. Because both N98 and B1 are interacting with the filament (see Sect.4.1),  we took the distance (4.44 kpc) of the filament as  those of the two H {\small II} regions. From Fig. 4, we measured a flux density of 630 mJy and 24 mJy at 1.4 GHz for the two H {\small II} regions, respectively. Finally,  we derived $N_{\rm Ly}\backsimeq$ 9.7$\times10^{47}$ ph s$^{-1}$ and 3.7$\times10^{46}$ ph s$^{-1}$ for the two H {\small II} regions.

Assuming an H {\small II} region expanding in a homogeneous medium,  the dynamical age was estimated by  \citep{Dyson80})
\begin{equation}
\mathit{t_{\rm  H {\small II}}}=\frac{4R_{\rm s}}{7c_{\rm s}}[(\frac{R_{\rm H {\small II}}}{R_{\rm s}})^{7/4}-1],
\end{equation}
where $c_{\rm s}$ is the isothermal sound speed of ionized gas, assumed to be 10 km s$^{-1}$, $R_{\rm H {\small II}}$ is the radius of the H {\small II} region, and $R_{\rm s}$ is the radius of the Str\"{o}mgren sphere given by $R_{\rm s}=(3N_{\rm Ly}/4\pi n^{2}_{i}\alpha_{B})^{1/3}$, where $N_{\rm Ly}$ is the ionizing luminosity, $n_{\rm i}$ is the initial H number density of the gas, and $\alpha_{B}=2.6\times10^{-13}$ cm$^{3}$ s$^{-1}$, which is the hydrogen recombination coefficient to all levels above the ground level.  We used the volume-averaged H$_{2}$ density (147.2 cm$^{-3}$) of the filament to determine $n_{\rm i}$ since the two H {\small II} regions  are located in or close to the filament. Adopting a radius of $\sim$1.6 pc and $\sim$0.6 pc for the two H {\small II} regions obtained from Fig. 4, we found that the  dynamical ages are 6.3$\times10^{5}$ yr and 3.9$\times10^{5}$ yr for the two H {\small II} regions associated with  N98 and B1, respectively.

Infrared bubbles were found to be associated with the filament. The bubbles will inject  kinetic energy into the filament. According to \citet{Lasker}, we estimate the kinetic energy of N98 and B1 by
\begin{equation}
\mathit{E_{\rm k}}=\frac{4}{3}\pi n_{i}m_{H}c^{2}_{s}R^{3/2}_{\rm s}[(\frac{7}{4}c_{s}R^{3/4}_{\rm s}t+R^{7/4}_{s})^{6/7}-R^{3/2}_{s}],
\end{equation}
where $m_{H}$ is the mass of a hydrogen atom, and $t$ represents the dynamical age of the H {\small II} region. As both N98 and B1  surround an H {\small II} region, we can estimate the ionization energy and thermal energy in the H {\small II} regions \citep{Freyer} as:
\begin{equation}
\mathit{E_{\rm i}}=\frac{4}{3}\pi n_{i}(\frac{7}{4}c_{s}R^{5/2}_{\rm s}t+R^{7/2}_{s})^{6/7}\chi_{0},
\end{equation}
\begin{equation}
\mathit{E_{\rm t}}=\frac{4}{3}\pi n_{i}(\frac{7}{4}c_{s}R^{5/2}_{\rm s}t+R^{7/2}_{s})^{6/7}kT_{\rm II},
\end{equation}
where $\chi_{0}$ is the ionization potential (13.6 eV) of hydrogen in the ground state, and $T_{\rm II}$ is the effective electron temperature (10$^{3}$ K) in the H {\small II} regions. The obtained results are listed in Table 3. From the Table 3, we can see that the ionization energy is one order of magnitude higher than kinetic energy and thermal energy.

\begin{table}[]
\begin{center}
\tabcolsep 1.5mm \caption{Energy components of H {\small II} regions. The first, second, and third columns are  the kinetic energy, ionization energy and thermal energy, respectively. The fourth column is the dynamical age.}
\def\temptablewidth{6\textwidth}
\begin{tabular}{lcccc}
\hline\hline\noalign{\smallskip}
Name  & $E_{k}$ & $E_{i}$ & $E_{t}$ & $t_{\rm HII}$ \\
      &($\times10^{47}$ergs)& ($\times10^{47}$ergs)& ($\times10^{47}$ergs)&  ($\times10^{5}$yr)  \\
  \hline\noalign{\smallskip}
N98&    2.2$\pm$0.7&    31.6$\pm$0.9 &  2.0$\pm$0.6  & 6.3\\
B1 &    0.1$\pm$0.03&    2.0$\pm$0.5 &  0.1$\pm$0.04 & 3.9\\
\noalign{\smallskip}\hline
\end{tabular}\end{center}
\end{table}

\subsection{Distribution of young stellar objects}
To investigate the star formation activity in the filament,  we use the GLIMPSE I Spring'07 catalog (highly reliable) to select young stellar objects (YSOs). Using a relation of $A_{v}=5.34\times10^{-22}N(\rm H_{2})$ \citep{Deharveng2009} and adopting a mean column density of 6.2$\times10^{21}$ cm$^{-2}$, we obtain an extinction of 3.3 mag in filament G47.06+0.26.  The IRAC $[5.8]-[8.0]$ versus $[3.6]-[4.5]$ color-color (CC) diagram  is a useful tool for identifying YSOs with  infrared excess \citep{Allen}. The effects of extinction are relatively small at these wavelengths.   Using the criteria of \citet{Allen}, we selected near-infrared sources with 3.6, 4.5, 5.8, and 8.0 $\mu$m detections in our observed region from the catalogue.   Figure 7 shows the $[5.8]-[8.0]$ versus $[3.6]-[4.5]$ color-color (CC) diagram, which can be used to classify  near-infrared sources into three regions: class I YSOs are protostars with circumstellar envelopes, class II YSOs are disk-dominated objects, and other sources. Class I YSOs have a timescale of the order of $\sim10^{5}$ yr, while the age of class II YSOs is $\sim10^{6}$ yr \citep{Andre94}.  Based on this criteria, we selected 75 Class I YSOs and 179 class II YSOs.

\begin{figure}[]
\vspace{0mm}
\includegraphics[angle=0,scale=0.3]{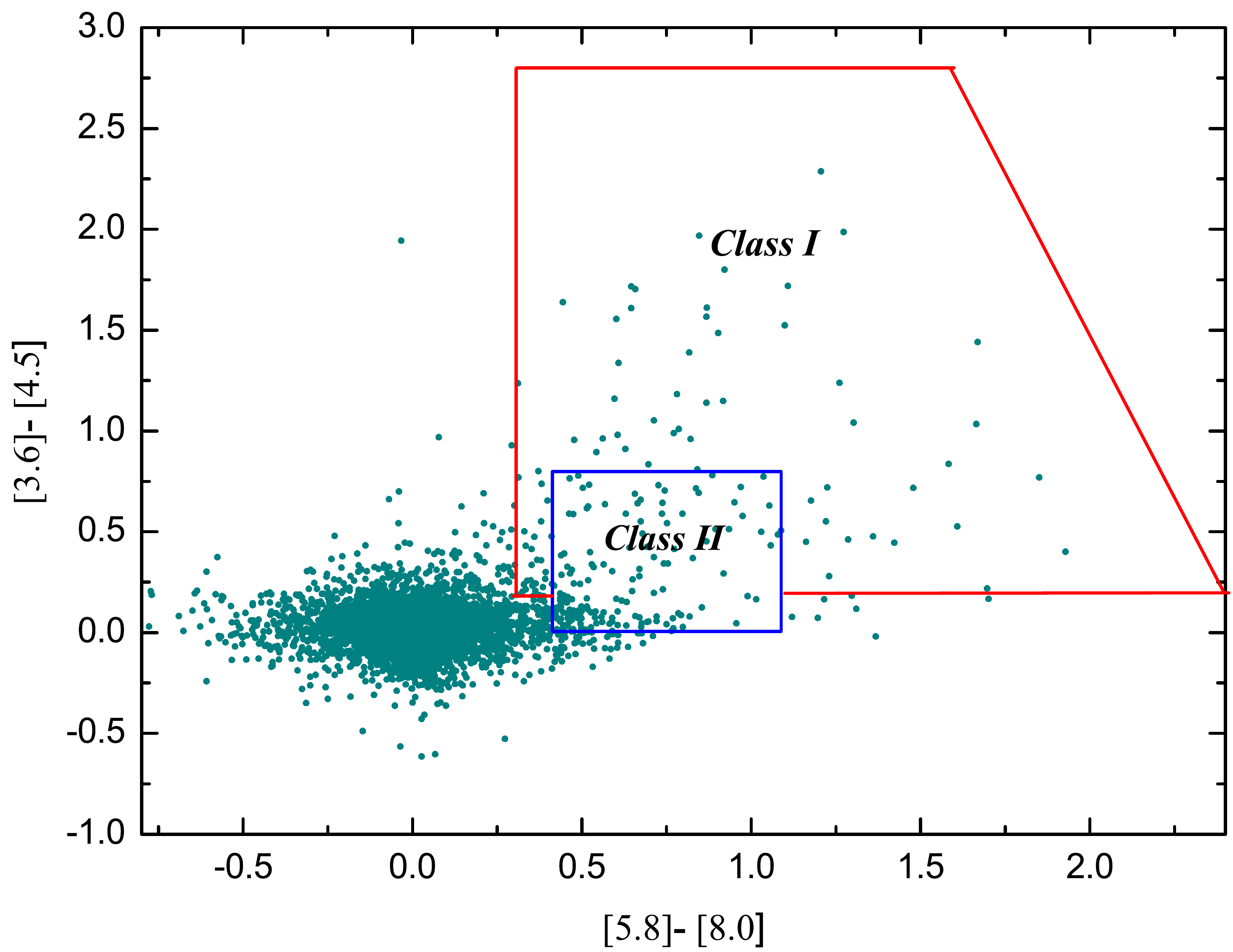}
\vspace{0mm}\caption{GLIMPSE {\bf color--color} diagram [5.8]--[8.0] versus [3.6]--[4.5]
for  all the selected sources in filament G47.06+0.26. Class I and II regions are indicated according to criteria given by Allen et al. (2004). Class I sources are protostars with
circumstellar envelopes and Class II are disk-dominated objects.}
\end{figure}

\begin{figure}[]
\vspace{0mm}
\includegraphics[angle=0,scale=0.315]{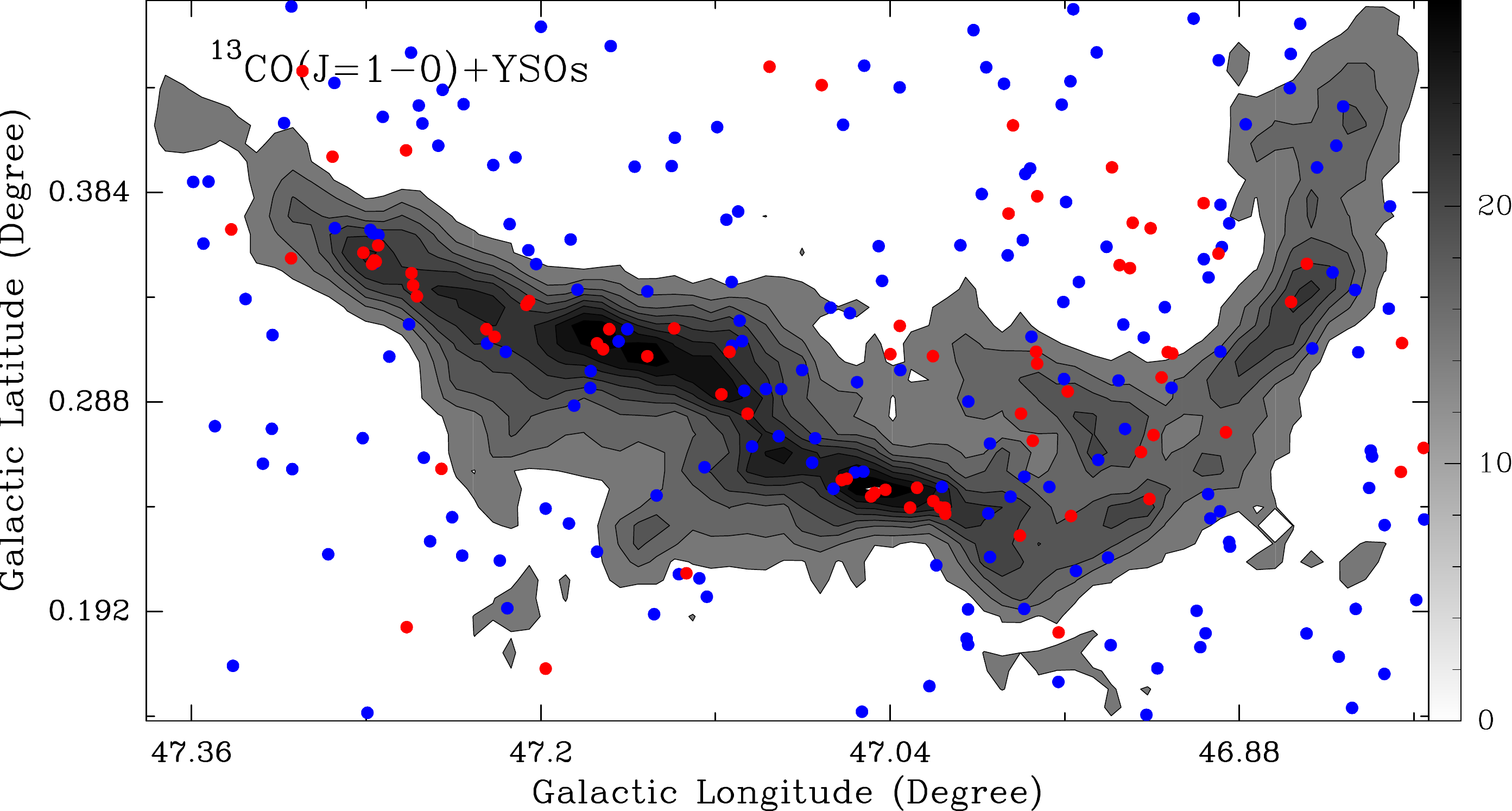}
\vspace{0mm}\caption{Positions of  class I and II  YSOs relative to the
$^{13}$CO ($J=1-0$) emission ({\bf gray} scale). Class I
YSOs are {\bf labeled} as red dots, and class II YSOs as  blue  dots.}
\end{figure}

The spatial distribution of these YSOs is shown in Fig. 8. To display the positional relation of YSOs with the filament, we also overlaid the selected class I and class II YSOs on the $^{13}$CO ($J=1-0$) emission (gray scale). We note that class II YSOs (blue dots) appear to be dispersedly distributed across the whole selected region. Class I sources (red dots) are found to be evenly along regions of high column density, which are mostly concentrated in the filamentary molecular cloud. In addition, region I of the filament has a higher concentration of class I YSOs than in the region II. To further confirm above  the results, we calculated the surface densities of the class I and class II YSOs inside and outside of the filament.  For the filament, the surface density of class I YSOs is 0.12 pc$^{-2}$. The region outside the filament has a surface density of 0.04 pc$^{-2}$, which represents the surface density of the total foreground and background class I YSOs. Getting rid of the total foreground and background objects, then we obtained a surface density of 0.08 pc$^{-2}$ for the filament. Because we only selected the near-infrared sources with the 3.6, 4.5, 5.8, and 8.0 $\mu$m emission, a surface density of 0.08 pc$^{-2}$ may be a lower limit.  Moreover, the filament has a surface density of 0.17 pc$^{-2}$ for class II YSOs, which is approximately equal to the value (0.15 pc$^{-2}$) outside of the filament, indicating that the class II YSOs are evenly distributed along the whole region.

\begin{figure}[]
\vspace{0mm}
\includegraphics[angle=0,scale=0.52]{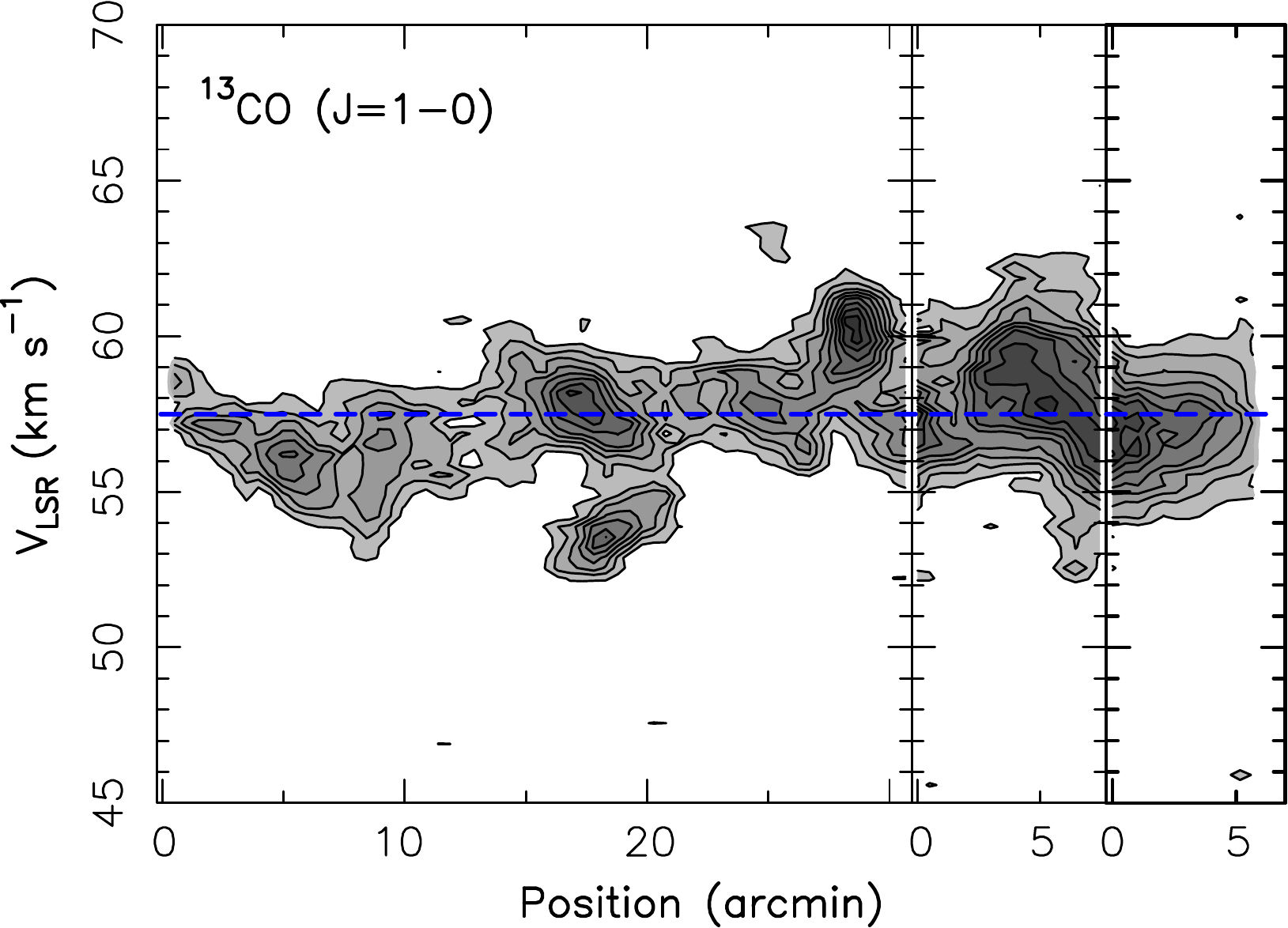}
\vspace{0mm}\caption{{\bf Position-velocity} diagrams of the $^{13}$CO ($J$=1-0)emission along filament G47.06+0.26 (see the long dashed lines in the right panel of Figure 2). The blue dashed line marks an LSR velocity of 57.5 km s$^{-1}$.}
\end{figure}

\begin{figure*}[]
\vspace{0mm}
\includegraphics[angle=0,scale=0.6]{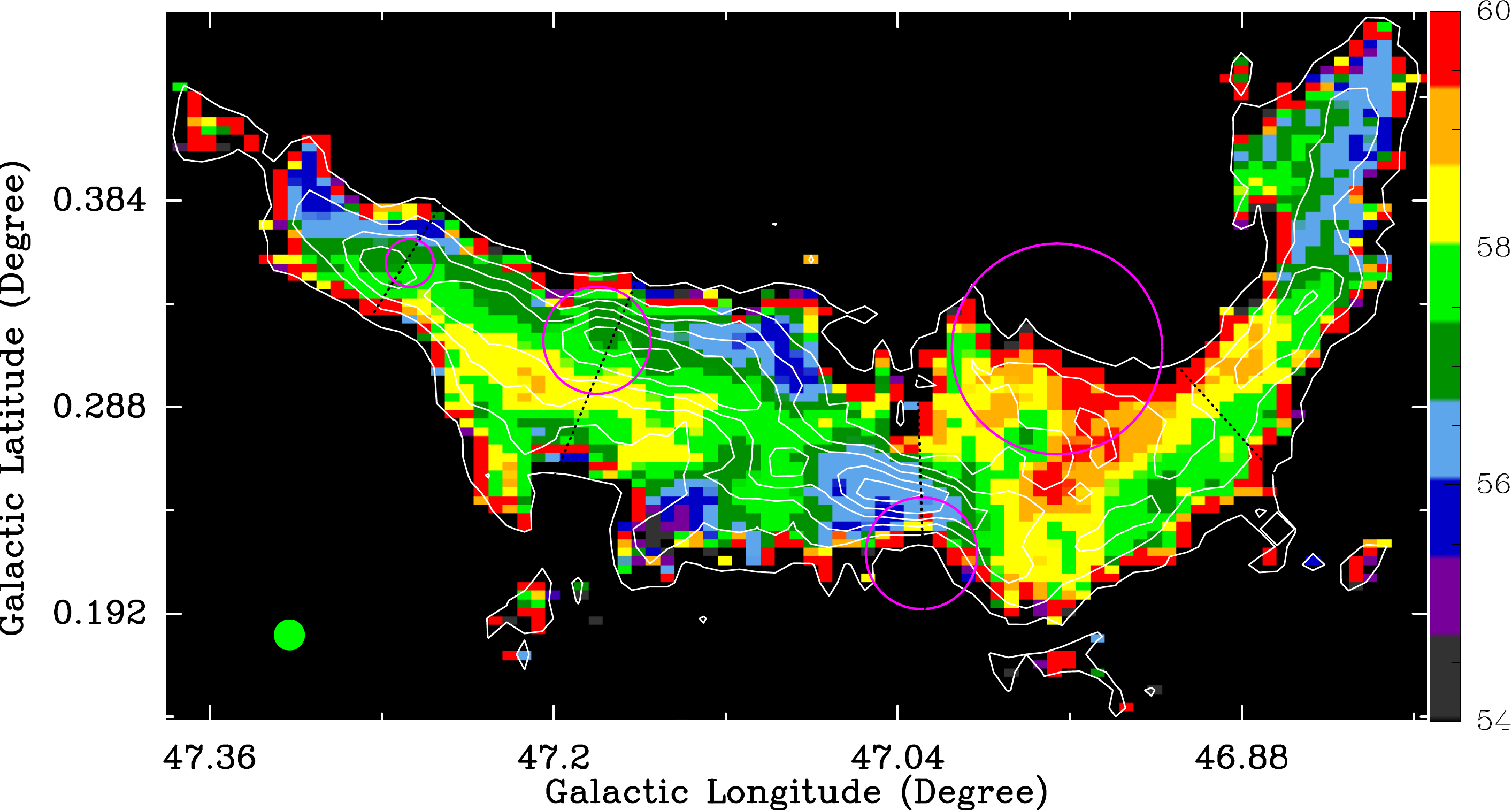}
\vspace{0mm}\caption{Velocity-field (moment 1) map of $^{13}$CO ($J$=1-0) overlaid with its integrated intensity contours (white). The four pink circles represent the identified infrared bubbles. The beam size is shown in  the lower-left corner. The right color bar is in units of km s$^{-1}$. The four black dashed lines show the velocity gradients perpendicular to filament G47.06+0.26. }
\end{figure*}

\section{Discussions}
\subsection{Dynamic structure of  the filament}
From the $^{12}$CO ($J$=1-0) and $^{13}$CO ($J$=1-0) emission, G47.06+0.26 shows a filamentary structure. The region traced by the $^{12}$CO ($J$=1-0) emission is more diffuse than that of $^{13}$CO ($J$=1-0) in G47.06+0.26. The C$^{18}$O ($J$=1-0) emission is weaker, which is consistent with the ATLASGAL 870 $\mu$m emission.  The filamentary structure can be divided into two regions (region I and region II), as shown in Fig. 2. We do not detect C$^{18}$O ($J$=1-0) and 870 $\mu$m emission in region II. The C$^{18}$O ($J$=1-0) and  870 $\mu$m emission likely traces the densest regions of the filament. Figure 9 shows a position-velocity (PV) diagram constructed from the $^{13}$CO ($J$=1-0) emission along the long filament G47.06+0.26. From this figure, we found that the filament not only  shows a single coherent object, but also contains several $^{13}$CO ($J$=1-0) clumps. Generally, filamentary molecular clouds harbor over-densities along their length \citep{Contreras}.
Recently several authors have identified the signatures of gas flows along filaments \citep[e.g.,][]{Peretto,Kirk, Tackenberg,zhang}. Because gas can be funneled along filaments and feed star-forming regions,  filamentary flows may play a key role in star formation. In numerical simulations, and  in comparison with cores in other regions, cores that are embedded within filaments have been shown to form more massive stars by accretion along filamentary flows \citep{Smith11}. Figure 10 presents a velocity-field (Moment 1) map  of $^{13}$CO ($J$=1-0) of G47.06+0.26 overlaid with its integrated intensity contours. From the velocity-field map, we can discern a  velocity gradient perpendicular to filament G47.06+0.26, but not along the filament.  \citet{Fernandez} suggested that such velocity-gradients, perpendicular to filaments may be due to motions associated with the formation and growth of the filaments within magnetized and turbulent sheet-like structures. As suggested by \citet{Smith16}, the formation of filaments is a natural consequence of turbulent cascades in the complex multiphase interstellar medium. Here we can estimate the turbulent energy of the filament G47.06+0.26, which is given by
\begin{equation}
\mathit{E_{\rm turb}}=\frac{1}{2}M\sigma^{2}_{3d},
\end{equation}
where $\sigma_{3d}$$\approx$$\sqrt{3}$$\sigma_{v}$, which is the three-dimensional turbulent velocity dispersion, $\sigma_{v}$ is $\triangle V_{\rm FWHM}/(2\sqrt{2\rm ln2})$, and $\triangle V_{\rm FWHM}$ is the mean full width at half-maximum (FWHM) of the $^{13}$CO $J$=1-0 emission. In the filament, we found the mean FWHM to be 3.8 km $\rm s^{-1}$. The mean thermal broadening can be given by $\Delta V_{\rm therm}$ =2.355$\times\sqrt{k T_{\rm ex}/(\mu_{g}m(\rm H_{2})) }$.  Adopting an excitation temperature ($T_{\rm ex}$) of 4.4-17.0 K, we derived  $\Delta V_{\rm therm}$ $\approx$ 0.3-0.5 km s$\rm ^{-1}$, which is much less than the mean FWHM of 3.8 km $\rm s^{-1}$ measured for $^{13}$CO ($J$=1-0). Hence, we did not consider the velocity dispersion to be caused by thermal broadening. Using a mass of $\sim$2.1$\times10^{4}\rm M_{ \odot}$ for the filament, we determined the turbulent energy of the filament to be about 1.6$\times10^{48}$erg.

Massive stars are a possible mechanism for driving turbulence through feedback from outflows, UV radiation fields, stellar winds, and supernova explosions in molecular clouds. Molecular outflows and supernova explosions have not been detected in G47.06+0.26. Instead, we found four infrared bubbles (N97, N98, B1, and  B2) in this region. N98 shows a comet-like structure at 8 $\mu$m. The H {\small II} region G047.028+0.232 is associated with N98  \citep{Anderson11}. The RRL velocity of G047.028+0.232 is roughly consistent with that of G47.06+0.26. Moreover, CO molecular gas of G47.06+0.26 shows an arc-like structure with a very intense emission towards N98. Hence, we conclude that N98 is interacting with  filament G47.06+0.26. From the velocity-field map (Fig. 10), we found that the adjacent proximity of G47.06+0.26 to N97 and N98 is the likely source of the velocity gradient. Moreover, the direction of the velocity gradients are towards  N97 and N98. Hence, the identified infrared bubbles are likely to be part of the turbulent sources in filament G47.06+0.26. We estimated the kinetic energy, ionization energy, and  thermal energy  of the H {\small II} region associated with N98. Both the obtained kinetic energy and thermal energy are far  smaller than the turbulent energy (1.6$\times10^{48}$erg) of filament G47.06+0.26. Compared to the kinetic energy and thermal energy, the ionization energy (3.2$\times10^{48}$erg) may inject more energy to G47.06+0.26, and help maintain the turbulence.

\begin{figure}[]
\vspace{0mm}
\includegraphics[angle=0,scale=0.3]{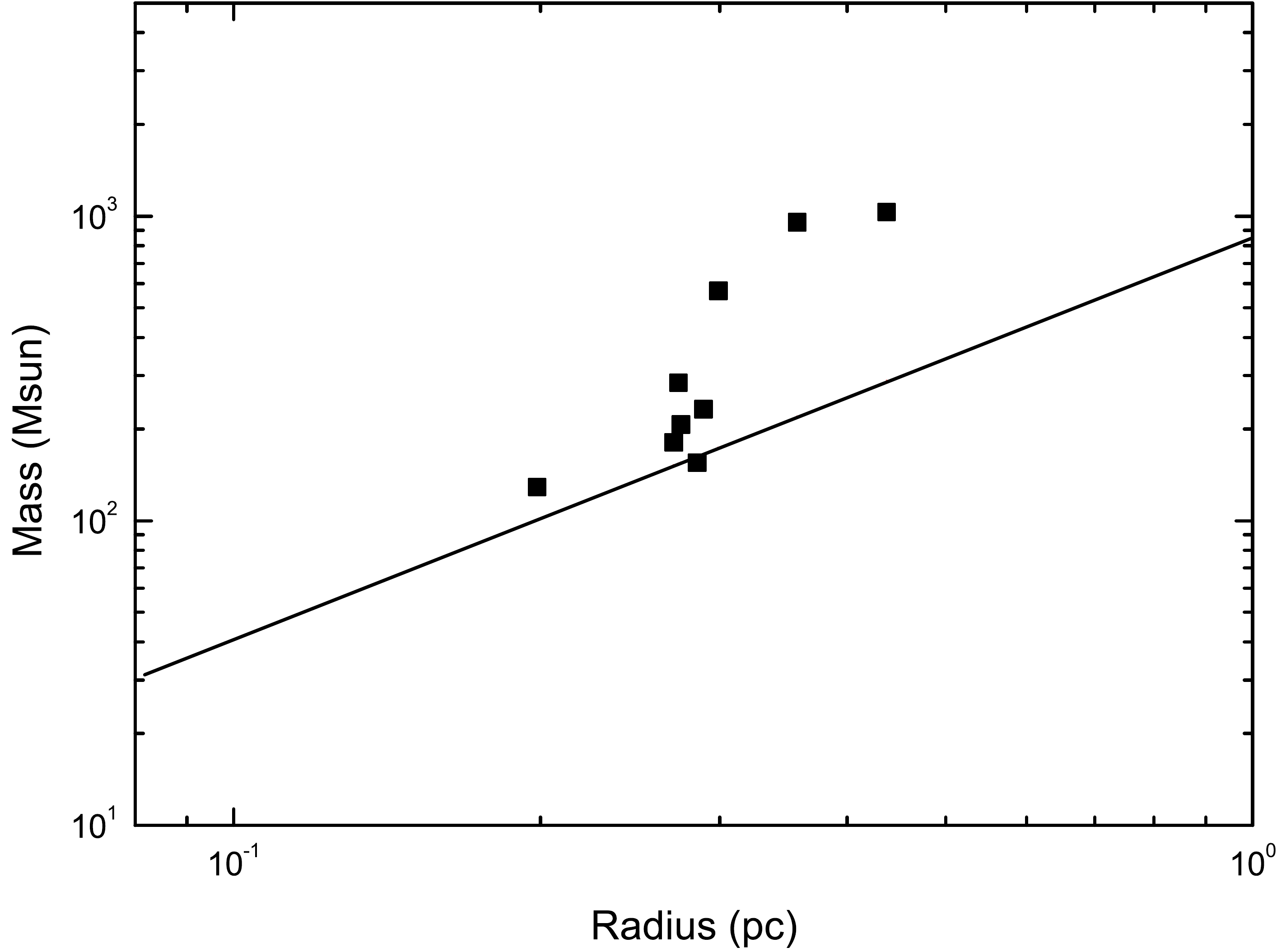}
\vspace{0mm}\caption{Mass versus radius for the BGPS sources embedded in the filament.  The black line delineates the threshold (Kauffmann \& Pillai 2010), which is used to separate the regimes under which massive stars can form (above the line) or not (below the line).}
\end{figure}

\subsection{Fragmentation and star formation}
Using the mass and length ($\sim$58.1 pc) of the filament, we derived a linear mass density $M/l$ of $\sim$361.5 $M_{\odot}$ $ \rm pc^{-1}$, which is an upper limit due to inclination and projection effects. If turbulence is dominant, a critical mass to length ratio can be estimated as ($M/l$)$_{\rm crit}$=84$(\Delta V)^{2}M_{\odot}$ $ \rm pc^{-1}$  \citep{Jackson}. Then, we obtain ($M/l$) $_{\rm crit}$ $\approx$ 1.2$\times$10$^{3}$ $M_{\odot}$ $ \rm pc^{-1}$, which is larger than $M/l$.  Hence, we conclude that external pressure (due to the  neighboring bubbles and H {\small II} regions) may help prevent the filament from dispersing under the effects of turbulence.

From the BGPS catalog,  we found ten BGPS sources in our observed region. In previous studies of BGPS sources, most show signs of active star formation \citep{Dunham,Xi}. We cannot  say if star formation is ongoing from our data. To determine whether the BGPS sources have sufficient mass to form massive stars, we must consider their sizes and mass. According to \citet{Kauffmann}, if the clump mass is $M(r)\geq870M_{\odot}(r/pc)^{1.33}$, then they can potentially form massive stars. Figure 11 presents a mass versus radius plot of the BGPS sources embedded in the filament. We find that all the BGPS sources lie at or above  the threshold, indicating that these clumps are dense and massive enough to potentially form massive stars. Nine out of the ten sources are associated with ATLASGAL 870 $\mu$m emission of G47.06+0.26. Although G47.06+0.26 is not gravitationally bound from the $^{13}$CO $J$=1-0 observation, the space distribution of the BGPS sources suggests that fragmentation may happen in the densest part of  the filament.  The total mass of filament G47.06+0.26 is $\sim$2.1$\times10^{4}\rm M_{ \odot}$, as derived from the $^{13}$CO $J$=1-0 observation.  These BGPS sources have a total mass of 3.8 $\times10^{3}\rm M_{ \odot}$. Hence, we can calculate a massive clump formation efficiency (CFE). The CFE is used to measure what fraction of molecular gas has been converted into dense clumps, which is given by $M_{\rm clump}$/$M_{\rm cloud}$ \citep{Eden2012}. The obtained CFE in the filament G47.06+0.26 is $\sim$18$\%$. In W3 giant molecular cloud (GMC), it is 26$\%$-37$\%$ in the compressed region by H {\small II} regions  and only 5$\%$-13$\%$ in the diffuse cloud \citep{Moore}. Moreover, \citet{Eden2013} used BGPS sources and Galactic Ring Survey data ($^{13}$CO) to check CFE changes with Galactic environment. They found that the total CFE values in the interarm and spiral-arm regions are about 4.9$\%$ and 5.5$\%$, but greater than 40$\%$ in the small-scale regions associated with  H {\small II} regions. The increases in the CFE  are related to local triggering due to feedback \citep{Eden2013} . Using the same data as \citep{Eden2012,Eden2013}, \citet{Battisti} derived that the mean CFE is about 11$\%$ in  the GMC  of the Milky Way.  Our obtained CFE value ($\sim$18$\%$) is greater than those in  the GMC  of the Milky Way, and in the interarm and spiral-arm regions. For the L1641 filamentary  molecular cloud in Orion, the CFE value is only 12$\%$  \citep{Polychroni}. The filament G47.06+0.26 is found to be associated with four infrared bubbles (N97, N98, B1, and  B2). We suggest that the  feedback from H {\small II} regions may be responsible for the higher CFE in the filament G47.06+0.26.

In addition, the bubbles can be produced by stellar wind and overpressure via ionization and heating by stellar UV radiation from O- and early-B stars. The presence of the infrared bubbles, suggests that several massive stars have formed in or near to the filament. The ionizing stars  in these bubbles may be the first generation of stars that form in such a filament. The formation of the bubbles can impact their environments, and in particular they  can  trigger the next generation of star formation \citep{Elmegreen}. We concluded that N98 is interacting with filament G47.06+0.26. From the GLIMPSE I catalog, we selected class I YSOs with ages of $\sim10^{5}$ yr on the basis of their infrared color indices. It is interesting to note that class I YSOs (red dots) are found to be distributed along regions of high column density, which are mostly concentrated in the filamentary molecular cloud. Moreover, we also find that the region adjacent to N98 contains more class I YSOs. The estimated dynamical ages are 6.3$\times10^{5}$ yr and 3.9$\times10^{5}$ yr for the two H {\small II} regions associated with N98 and B1, which are slightly greater than that of class I YSOs.  A possible picture would be that the expansion of the H {\small II} regions then compressed the filament, and triggered the formation of a new generation of stars, including what is now seen as class I YSOs in filament G47.06+0.26.

\section{Conclusions}
We present molecular $^{12}$CO ($J$=1-0), $^{13}$CO ($J$=1-0) and C$^{18}$O ($J$=1-0), infrared, and  radio continuum observations of filament G47.06+0.26. Molecular observations were obtained with the PMO 13.7 m radio telescope to investigate the detailed distribution of molecular material in G47.06+0.26. Our main results are summarized as follows:
 \begin{enumerate}
      \item
From the $^{12}$CO ($J$=1-0) and $^{13}$CO ($J$=1-0) emission, G47.06+0.26 shows a filamentary structure. The $^{12}$CO ($J$=1-0) emission is more diffuse than that of the $^{13}$CO ($J$=1-0) in G47.06+0.26. The C$^{18}$O ($J$=1-0) emission is weaker, which is consistent with the ATLASGAL 870 $\mu$m emission. The filament extends by about 45.0$^{\prime}$ along the east-west direction, whose mean width is about 6.8 pc, as traced by the $^{13}$CO ($J$=1-0) emission. The mass and mean number density of the filament are $\sim$2.1$\times10^{4}\rm M_{ \odot}$ and 147.2 cm$^{-3}$, respectively.
      \item
In the filament,  we obtained a linear mass density of $\sim$361.5 $M_{\odot}$ $\rm pc^{-1}$, which is smaller than its critical mass to length ratio, suggesting that external pressure (due to the neighboring bubbles and H {\small II} regions) may help prevent the filament from dispersing under the effects of turbulence. The velocity-field (Moment 1) map shows a clear velocity gradient perpendicular to filament G47.06+0.26, but not along the filament. Such velocity-gradients perpendicular to filaments may stem from the filament-formation process within magnetized and turbulent structures.
      \item
We determined that the turbulent energy of the filament is about 6.5$\times10^{48}$erg. Four infrared bubbles are found to be located in, or adjacent to, G47.06+0.26 along the line of sight. Particularly, the infrared bubble N98 shows a cometary structure and is expanding into G47.06+0.26. N98 is associated with H {\small II} region G047.028+0.232, with an age of 6.3$\times10^{5}$ yr. We estimated the kinetic energy, ionization energy, and  thermal energy  of H {\small II} G047.028+0.232. Both the obtained kinetic energy and thermal energy are far smaller than the turbulent energy (1.6$\times10^{48}$erg) of filament G47.06+0.26. Compared to the kinetic energy and thermal energy, the ionization energy (3.2$\times10^{48}$erg) may inject more energy into the G47.06+0.26, and could help maintain the turbulence.
      \item
We found nine BGPS sources within G47.06+0.26. These BGPS sources have sufficient mass to form  massive stars. The obtained clump formation efficiency (CFE) is  $\sim$18$\%$  in the filament G47.06+0.26, which is greater than the value ($\sim$11$\%$) in  the GMC  of the Milky Way. We suggest that the feedback from H {\small II} regions may be responsible for the higher CFE. The feedback from the H {\small II} regions may cause the formation of a new generation of stars and the higher CFE in filament G47.06+0.26. The selected class I YSOs are clustered along G47.06+0.26. Comparing the ages of the two H {\small II} regions associated with N98 and B1 with that of class I YSOs ($\sim$10$^{5}$ yr), we infer that the expansion of the H {\small II} regions may trigger the formation of a new generation of stars, including what is now seen as class I YSOs in filament G47.06+0.26.
\end{enumerate}

\begin{acknowledgements}
We are very grateful to the anonymous referee for his/her helpful comments and suggestions.
We are grateful to the staff at the Qinghai Station of PMO for their assistance during the observations. Thanks for the Key Laboratory for Radio Astronomy, CAS, for partly supporting the telescope operation. This work  has made use of data from the Spitzer Space Telescope, which is operated by the Jet Propulsion Laboratory, California Institute of Technology under a contract with NASA. The ATLASGAL project is a collaboration between the Max-Planck-Gesellschaft, the European Southern Observatory (ESO) and the Universidad de Chile. It includes projects E-181.C-0885, E-078.F-9040(A), M-079.C-9501(A), M-081.C-9501(A) plus Chilean data. This work was supported by the National Natural Science Foundation of China (Grant No. 11363004, 11403042, and 11673066).
\end{acknowledgements}

\bibliographystyle{aa}
\bibliography{references}
\end{document}